\documentclass[twocolumn,showpacs,preprintnumbers,amsmath,amssymb,10pt,a4paper]{revtex4}
\usepackage{graphicx}
\usepackage{epsfig}

\usepackage{amsmath}
\usepackage{amsfonts}
\usepackage{amssymb}
\usepackage{dcolumn}
\usepackage{bm}

\begin{document}

\title{Vortex core magnetization dynamics induced by thermal excitation}

\author{Tiago S. Machado$^{1}$}
\author{Tatiana G. Rappoport$^{2}$}
\author{Luiz C. Sampaio$^{1}$}%


\affiliation{$^{1}$Centro Brasileiro de Pesquisas F\'{\i}sicas,
  Xavier Sigaud, 150, Rio de Janeiro, RJ, 22.290-180, Brazil
}%

\affiliation{$^{2}$Instituto de F\'{\i}sica, Universidade Federal
  Fluminense, Rio de Janeiro, RJ, 24.210-346, Brazil
}%


\date{\today}

\begin{abstract}
We investigate the effect of temperature on the dynamic properties of magnetic vortices in small disks. Our calculations use a stochastic version of the Landau-Lifshitz-Gilbert (LLG) equation, valid for finite temperatures well below the Curie critical temperature. We show that a finite temperature induces a vortex precession around the center of the disk, even in the absence of other excitation sources. We discuss the origin and implications of the appearance of this new dynamics. We also show that a temperature gradient plays a role similar to that of a small constant magnetic field. 
\end{abstract}

\maketitle


The control and manipulation of the magnetization in magnetic materials is one of the most interesting challenges in the field. In the last 10 years it has been shown that spin polarized current~\cite{spintorque1,spintorque2}, 
light ˜\cite{light}, and electric field \cite{ohno2000} can be used to modify a magnetic state, 
providing new opportunities for technological applications.  More recently, it has also been observed that heat can act as an excitation source, 
producing a change in the magnetization. When different temperatures are applied at opposite ends of a magnetic material, leading to a temperature gradient, 
a pure spin current is generated, an effect known as the spin Seebeck effect (SSE)~\cite{uchida2008}. Numerical simulations also indicate that a temperature gradient can move domain walls in nanowires \cite{nowak2011}. 

Depending on their length and thickness, microsized disks 
made by magnetic materials like Permalloy (Py) can exhibit a vortex in its center~\cite{shinjo,cownburn1}.
Under excitation of an in-plane magnetic field or a spin polarized 
current in the form of short pulses or an a.c. resonant excitation, 
the vortex core moves around the center of the disk, and 
depending on the excitation intensity, the core magnetization can be 
 reversed.

In this paper we investigate the effect of the temperature 
on the vortex core magnetization dynamics in Py disks. 
We show that even in the absence of a temperature gradient, heat can induce 
dynamics in the system. The vortex core rotates around the center of the disk and the amplitude of the trajectory depends on the temperature. We discuss the consequences of this effect on the hysteresis curve and also address the effect of a temperature gradient in the dynamical process.  

In order to simulate the vortex core magnetization dynamics
we used the Landau-Lifshitz-Gilbert (LLG) equation. 
The effect of the temperature is introduced  in the calculations by including 
a stochastic term in the total field~\cite{stoch}.  The new stochastic term 
mimics random fluctuations induced by the interaction of the nanomagnet
with a thermal bath.

The LLG equation is then given by
\begin{equation}
  (1+\alpha^{2})\dfrac{d}  {d\tau}\textbf{m}(t)=-\gamma_{0}\textbf{m}
  (t)\times\textbf{h}-\alpha\textbf{m}(t)\times[\textbf{m}(t)
  \times\textbf{h}],
  \label{llg}
\end{equation}
where $\alpha$ is the Gilbert damping constant, $\gamma_{0}$ 
is the gyromagnetic ratio, \textbf{m} is the normalized magnetization
of a cell, 
and \textbf{h}$=$\textbf{h}$_{eff}$+$\nu$\textbf{h}$_{r}$, 
\textbf{h}$_{eff}$ being the effective field and 
$\nu$\textbf{h}$_{r}$ the noise field. 
\textbf{h}$_{eff}$ contains the exchange and  
dipole-dipole interactions, and the applied magnetic field.

The noise term has the form of a random magnetic torque 
$-\nu$\textbf{m}$\times$\textbf{h}$_{r}$~\cite{stoch}, where \textbf{h}$_{r}$ is
the random vector whose components are independent, 
and $\nu^2={2\alpha k_{B}T}/{\mu_{0}M_{s}^2 V}$ is a parameter 
which measures the intensity of thermal noise;
$k_{B}$ is the Boltzmann constant, $T$ is the temperature, 
$\mu_{0}$ is the magnetic permeability, $M_{s}$ 
is the magnetization saturation and $V$ is the volume of the cell. 
It is important to note that this approach is valid for temperatures well below the critical temperature
of the magnet.  In this regime the exchange interaction preserves the magnetization 
uniformity inside each cell despite of the thermal perturbation.  As a consequence, 
the random thermal torque conserves the magnetization magnitude 
in the stochastic dynamics. 

\begin{figure}
    \includegraphics[width=0.85\columnwidth]{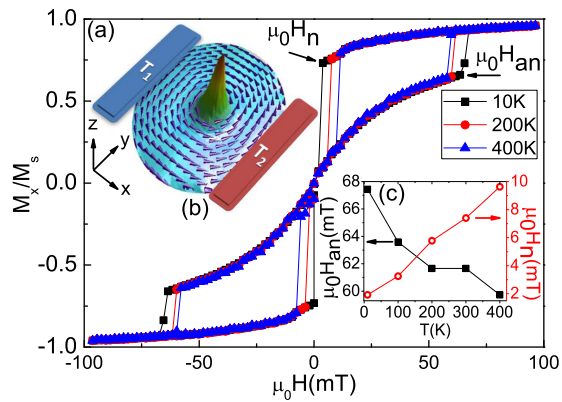}
    \caption{{ (a) Permalloy disk with electrical contact pads at 
    temperatures $T_{1}$ and $T_{2}$ where
    $T_{2} =T_{1}$+$\Delta T$. (b) m$_x$ as a function of an in-plane magnetic field $h_x$ for disks at different temperatures  $T=T_1=T_2$. (c) Nucleation $h_{n}$ and annihilation $h_{an}$ fields as a function of temperature.}
    \label{fig1}}
\end{figure}

Figure~\ref{fig1}(b) illustrates our setup. We consider a Py disk exhibiting a vortex and we connect two 
thermal contacts at opposite sides of the disk. The thermal contacts 
can in principle be at two different temperatures $T_{1}$ and $T_{2}$. 
In order to calculate the temperature distribution inside the disk, we used a relaxation method 
to solve the Laplace equation, as was done previously to obtain the voltage drop for the same geometry~\cite{jap}. 
 
The disk has a diameter 
of 300 nm and thickness of 20 nm, and was discretized in cells of  
$5\times 5 \times 5$ nm$^3$. The parameters associated with 
the LLG equation are the saturation magnetization 
$M_s=8.6 \times 10^5$ A/m, the exchange coupling $A=1.3\times10^{-11}$ 
J/m and the Gilbert damping constant $\alpha=0.01$. 
Given the temperature of the thermal contacts,
we solve the LLG equation numerically using the fourth-order 
Runge-Kutta approximation with a code we have written for this purpose.

Let us initially consider the disk with $T=0$ in both contacts, which  
corresponds to the usual LLG equation. As it is expected, we 
find a magnetic vortex structure with the vortex 
core at the center of the disk, as shown in Figure~\ref{fig1}(b). 
We now consider the two contacts at same temperature $T\neq0$ and calculate the hysteresis curve 
for an in-plane external magnetic field (see Fig.~\ref{fig1}(a)). 
When applying a magnetic field, the vortex
core moves towards the disk edge. Its expulsion occurs at a
critical annihilation field $h_{an}$, leading 
the disk to a mono-domain state.  For decreasing fields, the vortex nucleates again at a lower nucleation field $h_{n}$ where a sharp transition is observed from the uniform state to the vortex state. One can observe that both ${h}_{an}$  and  $h_{n}$ are temperature dependent. However, while ${h}_{an}$ decreases monotonically with temperature,  $h_{n}$  increases with $T$ ( Fig.~\ref{fig1} c). This result is consistent with experiments~\cite{vortexhyst} for intermediate $T$, which corroborate the way our model deals with temperature. 
\begin{figure}
    \includegraphics[width=0.95\columnwidth]{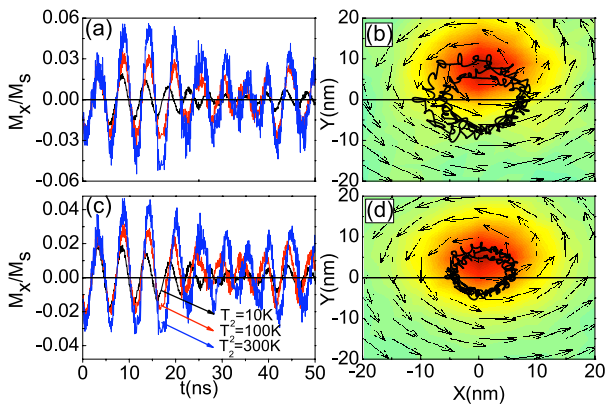}
   \caption{ Time evolution of the m$_{x}$ component of the 
   disk at different temperatures: (a) $\Delta T=0$ and (c)  $\Delta T \neq 0$. Snapshot of the magnetic configuration of the disk where the arrows represent the in-plane magnetization and the colors represent out-of-plane magnetization. The black line represents the core trajectory: (b) $\Delta T=0$ and (d)  $\Delta T \neq 0$. } \label{fig2}
\end{figure}

An interesting picture emerges when we calculate the time evolution of the vortex core magnetization 
at a finite temperature ($T \neq 0$). Surprisingly, the thermal fluctuations play a role similar to an external excitation such as magnetic field or spin polarized current.  In Fig. \ref{fig2}(b), we can observe that the thermal noise displaces the vortex 
core from the center of the disk. The vortex precesses around it, producing an orbital
trajectory. The precession frequency is the natural oscillatory frequency, here approximately 200 MHz. Figure~\ref{fig2}(a) shows the time variation of the $m_x$ 
component of the magnetization at 10, 100 and 300 K. One sees that $m_x$ has a sinusoidal dependence with time. Both amplitude and variance increase with temperature while the oscillating frequency remains constant.  
The system has a transient in which the amplitude increases with time, decreasing again later.

 We can use a simplified model to understand the role of the thermal fluctuations in the dynamics of the vortex core. Let us consider Thiele's formulation for the dynamics of a vortex core \cite{thiele}: $\textbf{G}\times {\dot{\textbf{r}}}+{\textbf{F}}^d+{\textbf{F}}^u=0$
where $\textbf{r}$ is the vortex's center and $\textbf{G}=-G\hat{\textbf{z}}$ is the gyroforce determined by the vortex non-uniform magnetization distribution~\cite{kim}. ${\textbf{F}}^d=-D\textbf{v}$ is the dissipation force and $\textbf{F}^u=-\vec{\nabla}U(r)$ where $U(r)$ is the magnetic potential that includes the exchange, anisotropy and magneto elastic energy.  In the linear approximation,  $U=1/2kr^2$ . In the presence of a random field $\nu\bf{h}_r$, which mimics thermal fluctuations, the equation has an extra force $\textbf{F}_r=-\mu(\hat{\textbf{z}}\times \nu{\textbf{h}})$, where $\mu=2/3\pi RhM_s c$, $R$ is the disk radius, $h$ is the disk thickness and $c$ is the vortex quirality~\cite{kim}.  It is possible to decouple the $\hat{x}$ and $\hat{y}$ components of the coupled first order differential equations \cite{kim} and the resulting  equations for the $x$ and $y$ components are equivalent to an underdamped harmonic oscillator in the presence of a random force: $\ddot{x}+2\gamma \dot{x}+\omega_0^2x=F(t)$.  This stochastic equation was first solved by Chandrasekhar~\cite{chand} for different initial conditions. In our case, the initial conditions are such that $\textbf{r}(t=0)=\textbf{v}(t=0)=0$. These initial conditions are also a  solution of Thiele's differential equation in the absence  of thermal fluctuations where both kinetic and potential energies are equal to zero. However, when these fluctuations are included, during a transient the system gains an extra energy $k_BT$ which reflects in an increase of both potential and kinetic energies. As a result, the vortex moves away from the disk center and acquires a velocity. The vortex then begins to precess around the disk center with its natural oscillatory frequency, with the initial precession radius proportional to the temperature (and hence, to the amplitude of the fluctuations). After the transient, the vortex has the behavior observed in Chandrasekhar's solution.  $\langle \textbf r \rangle$ follows an underdamped circular motion and the variance of the velocity increases with time, approaching a constant value proportional to the temperature.

For the LLG dynamics, even though the random term acts differently on each individual cell, we observe a behavior very similar to the one predicted by Thiele's equation. However, for high temperatures  we see a small shift of the frequency of the circular motion to values that are lower than the natural frequency of the system. A similar shift was observed in recent experiments~\cite{kamionka2011}. The discrepancy between the LLG dynamics and  Thiele's equation for high temperatures is expected, since the magnetic structure of the disk is modified in this temperature regime. 

In our simulations the oscillating radius is about 2-3 nm at room temperature. However, larger values of orbital radius
could be measured by heating the sample above 
room temperature. For instance, it is possible to use  techniques based 
on x-ray magnetic circular dichroism in time-resolved x-ray 
microscopy experiments to observe the trajectory. Recently, St\"{o}hr {\it et al.} have presented 
space- and time-resolved images of the magnetic vortex 
resonant movement in a spin-valve nanopillar ˜\cite{yu2011}. Driven by 
a dc spin current, the spin-torque effect gives the
vortex a resonant movement with a radius of $\sim$ 10 nm. 
They also measured the vortex position without spin 
current. This was done via independent measurements, which should show the 
static position of the vortex. However, the position 
is not static, following a trajectory with a radius of $\sim$ 2-3 nm. We suggest that this  could be related to the the thermal effect we observe in our simulations. 

Returning to the hysteresis, the decrease of ${h}_{an}$ 
with temperature can now be better understood. It occurs because 
in addition to the applied magnetic field that moves the vortex core 
from the center of the disk along the $y$ axis, the thermal excitation 
produces an extra shift of the vortex core. Since the radius 
increases with T, a smaller field is then 
needed to allow the vortex core to reach the disk edge leading 
to the vortex annihilation.   

It is known that a thermal gradient can be mapped onto an additional torque in the LLG equation~\cite{bauer}. So, the effect of the thermal gradient should be similar to that of an external field. On the other hand, here we have seen that the temperature itself also modifies the dynamics of the vortex core.   To investigate the influence of the thermal gradient on the 
dynamics, we keep one side of the magnet at $T=T_{1}$ with the opposite side at $T_{2}$. 
For $\Delta T$ varying from 25 to 75 K we found that 
the vortex core also oscillates at the natural 
frequency (200 MHz) and with increasing 
amplitudes (see Fig~\ref{fig2}d). However $m_{x}$
is shifted to positive values, meaning the orbit center is 
displaced from the center of the disk to the positive $y$-axis, 
which is perpendicular to the heat flow (see Fig~\ref{fig2}c). These data show that 
$\Delta T$ plays the same role of a small magnetic field applied in the $x$ 
direction, ${h}_{x}$. 

\begin{figure}
    \includegraphics[width=0.9\columnwidth,clip]{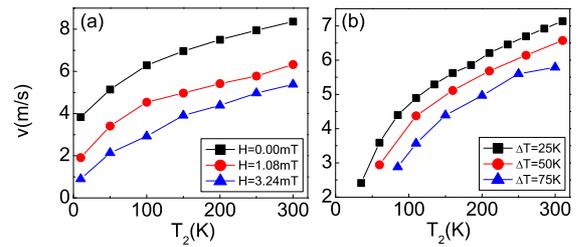}
   \caption{Vortex core speed in function of temperature
   with an applied magnetic field $h_x$and $\Delta T=0$ (a), and 
   without magnetic field and $\Delta T \neq 0$ (b).}
    \label{fig3}
\end{figure} 

To further investigate the similarities between the temperature gradient and a constant magnetic field,  we compared it with a system in the presence of an external magnetic field at a constant temperature $T\neq 0$.  We calculated the
vortex core speed considering the presence of \textbf{h}$_{x}$ or $\Delta T$ independently, as can be seen in  Fig~\ref{fig3}(a) and (b) respectively.  Both dynamics are very similar. For a fixed temperature $T_2$, the external field displaces the vortex core, increasing the potential energy and decreasing the kinetic energy of the core. This results in a decrease of the vortex speed (Fig~\ref{fig3}(a)). Similarly, for a fixed $T_2$ a decrease in $T_1$ by $\Delta T$ also results in a decrease of the kinetic energy and speed (Fig~\ref{fig3}(b))

In conclusion, we show that  for small disks a constant temperature can drive the system away from its equilibrium and produce a precession of the vortex core around the center of the disk.  As a result, temperature can be used, in conjunction with other external excitations, to manipulate the magnetic state in specific geometries.  

We would like to thank M. A. de Menezes for useful discussions. 
This work was supported by CNPq and FAPERJ. LCS 
acknowledges ``INCT de Fot\^onica'' for financial support.

\end{document}